\documentstyle[11pt,newpasp,twoside,epsf]{article}
\newcommand{\OmM}{\Omega_{\rm M}}
\newcommand{\hn}{\hat{n}}
\newcommand{\inu}{i_\nu}
\newcommand{\jnu}{j_\nu}
\newcommand{\sigT}{\sigma_T}
\newcommand{\fgas}{f_{\rm gas}}
\newcommand{\Da}{D_{\rm ang}}
\newcommand{\Mdet}{M_{\rm det}}
\newcommand{\Ynoise}{Y_{\rm noise}}
\newcommand{\Ompix}{\Omega_{\rm pix}}
\newcommand{\fwhm}{\theta_{\rm fwhm}}

\def\edcomment#1{\iffalse\marginpar{\raggedright\sl#1\/}\else\relax\fi}
\reversemarginpar

\begin{document}
\title{Searching and studying clusters with the SZ effect}
 \author{James G. Bartlett}
\affil{Observatoire Midi--Pyr{\'e}n{\'e}es, Toulouse, France\\
{\em New institution:} APC -- Universit{\'e}
   Paris 7, Paris, France}

\begin{abstract}
I discuss galaxy cluster surveys based on the Sunyaev--Zel'dovich
effect and their relevance for cosmological studies.
The unique aspects of cluster selection by this method
are emphasized and certain issues of surveying are addressed.
Finally, I briefly present prospects for upcoming surveys.
\end{abstract}

\section{Introduction}

     Galaxy cluster catalogs are useful for cosmological studies
for a number of reasons: cluster abundance, distribution and
evolution reflect the structure formation process, which
depends sensitively on certain cosmological parameters,
such as the matter density $\OmM$ (Oukbir \& Blanchard 1992, 1997; 
Bahcall 1998; Bartlett 1997; Borgani et al.
1999 \& these proceedings; Eke et al. 1998; Henry 1997; Viana \&
Liddle 
1999); and cluster internal
properties provide clues to the nature of dark matter and
on certain mechanisms associated with galaxy formation, 
such as heating and feedback. 

     The construction of well--defined catalogs appropriate 
for statistical studies demands good 
understanding of the cluster selection criteria (see Postman, these
proceedings).  
These criteria in turn depend on the type of observations
-- optical, X--ray, radio -- as well as on the data
analysis proceedures.  Each type of observation has its
own advantages and disadvantages that we may refer to
as its intrinsic biases.  For example, X--ray cluster
selection eliminates the problem of projection effects
suffered by optical observations, but on the other hand
is significantly influenced by the rather poorly understood
physics taking place in the core region. 

     The question that arises is, what kind of selection
is best?  The answer clearly depends on the utlimate 
use of the catalog, and for cosmological studies we may argue
that the most ideal selection would be based on cluster mass.
This is because cluster mass and redshift are the two primary
theoretical cluster descriptors.  From this line of argument,
understanding cluster selection criteria is reduced to 
relating the observational selection criteria to cluster
mass and redshift.  This is a useful viewpoint, for it
provides a framework for quantifying the biases 
associated with each type of observation for the purposes
of cosmological studies.  We do not argue that one particular
observational method is the best; rather, that the important
point is to use several methods with {\em different}
biases.  

     The Sunyaev--Zel'dovich (SZ) effect (Sunyaev \& Zel'dovich 1972)
will soon open the way to a new method of surveying for clusters.
Several ground--based instruments optimized for SZ surveying 
are currently under construction, and the Planck satellite will
supply an almost full--sky catalog of several $10^4$ clusters
at the turn of the decade.  In anticipation and in the spirit 
just outlined, we shall now examine in more detail 
this new kind of survey (the `SZ band'), and in particular
the nature of its cluster section.  

     This potential of surveying for clusters with the SZ effect
has been appreciated for some time.  To my knowledge, the first to 
give source count predictions were Korolyov, Sunyaev, 
\& Yakubstev (1986), and many authors have since considered the 
issue.  An indicative, useful, though by no means exhaustive, 
list of references
focused on SZ surveying (i.e, counts and redshift distributions
of SZ detected clusters)
would include: Bond \& Meyers (1991), Bartlett \& Silk (1994), Markevitch
et al. (1994), Barbosa et al. (1996), Eke, Cole, \& Frenk (1996), 
Aghanim et al. (1997), Colafrancesco et al. (1997), da Silva et
al. (2000), Holder et al. (2000), Lo et al. (2000), Benson, Reichardt,
\& Kamionkowski (2001), Fan \& Chuieh (2001), Kay, Liddle, \& 
Thomas (2001), Kneissl et al. (2001), Springel, White, \& Hernquist (2001),
Xue \& Wu (2001)\footnote{The reader will appreciate the 
precarious position of the author in giving such a list. 
In offering my apologies to those that I have unjustly, but 
without malice, forgotten, I ask no more than to be corrected
by a kind email in reminder.}.

\section{SZ Cluster Selection}

     In the non--relativistic limit (which applies in 
practice to all but the hottest clusters), the SZ effect
may be written as 
\begin{eqnarray*}
\delta \inu(\hn) & = & y(\hn) \jnu \\
y                & = & \int_{los} \frac{kT}{mc^2}n\sigT dl 
\end{eqnarray*}
where $\delta \inu$ is the brightness change induced in the
cosmic microwave background (CMB) by compton scattering
off electrons (mass $m$) in the hot intracluster medium (ICM) 
at temperature $T$ and electron density $n$. The Compton $y$--parameter
measures the amplitude of the energy transfer
to the photons ($\sigT$ is the Thompson cross section);
and the function $\jnu$ describes the spectral form
of the distortion.  It is convenient to consider the
effect integrated over the entire cluster face:
\begin{equation}
Y = \int d\Omega\; y(\hn) \propto \fgas\frac{M\langle T\rangle}{\Da^2}
\end{equation}
where now $M$ is the cluster (virial) mass, $\fgas$ is the ICM 
gas mass fraction, $\langle T\rangle$ is the (true) mean electron
temperature (i.e., particle weighted), and $\Da(z)$ is the
angular--size distance.  Notice that $Y$ is measured in, for 
instance, square arcmins.

     It is from these equations that we begin to understand 
the nature of SZ cluster selection (or detection).  Let us emphasize
four aspects:
\begin{enumerate}
\item The spectral signature $\jnu$ is unique -- negative
  at low frequency, positive at high frequency, with a fixed
  zero point at $\lambda\sim 1.4$~mm -- and the same for 
  all clusters ($\jnu$ depends only
  on the observation frequency).  There is therefore no k--correction;
\item The integrated effect $\propto \Da^{-2}$, which is 
  the statement that the surface brightness is independent
  of redshift $z$.  This is in contrast to other emission
  mechanisms, that vary as the luminosity distance $D_{\rm lum} =
  (1+z)^2\Da$, and which suffer from `cosmic dimming' (the redshift);
\item The integrated signal is directly proportional to 
  the total ICM thermal energy $\fgas M\langle T\rangle$
  (this is essentially the definition of the mean temperature
  $\langle T\rangle$).  We may therefore
  expect a tight correlation between the measurable, $Y$ and
  the cluster mass, e.g., $Y\propto \fgas M^{5/3}$, especially
  in the more massive systems where non--graviational effects
  appear less important. 
\item Cluster detection and subsequent modeling may be based
  on the same measurable quantity $Y$.
\end{enumerate}

     A particularily useful quantity for characterizing
the selection properties of a survey is the minimum
detectable mass as a function of redshift, $\Mdet(z)$.
For example, it determines the source counts as
\begin{equation}\label{eq:counts}
\frac{dN}{d\Omega} = \int_0^\infty dz\; \frac{dV}{dzd\Omega}
   \; \int_{\Mdet(z)}^\infty dM\; \frac{dn}{dM}(M,z)
\end{equation}
where $dn/dM$ is the mass function.  A comparison of 
$\Mdet$ from all--sky X--ray and SZ (the Planck curve)
surveys is made in Figure 1.  The X--ray survey flux limit 
corresponds to that of the RASS.  An interferometer--like
similar to those proposed (see below) and capable of 
surveying several square degrees from the ground is also shown.
To make the figure,
a non--evolving $L_x-T$ relation of the form given by
Arnaud \& Evrard (1999) and a $T-M$ relation from 
Evrard, Metzler, \& Navarro (1996) have been adopted.  The cosmological
model is an open, low--density model, and 
Eq. (1) was used with a constant 
gas mass fraction $\fgas=0.06 h^{-1.5}$ 
(Evrard 1997)

\begin{figure}
\plotone{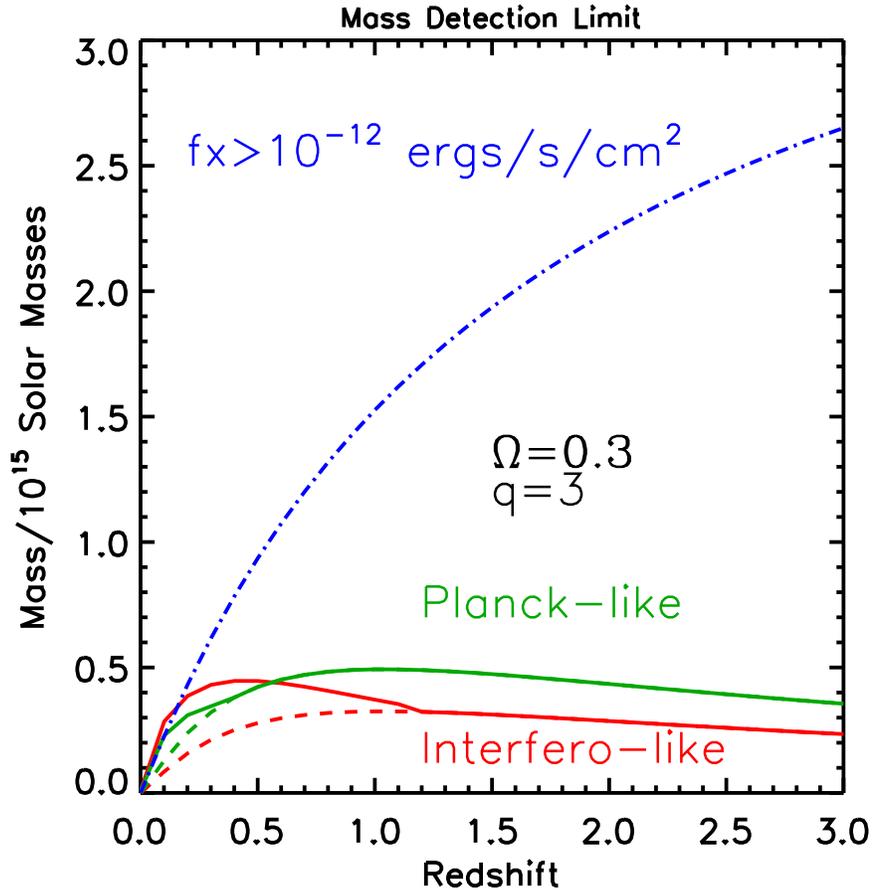}
\caption{Minimal detectable mass as a function of redshift
for a Planck--like all--sky survey, an interferometer--like 
(limited sky, ground--based) survey
and for a flux--limited X--ray survey similar to the RASS.
For the SZ curves, the dashed lines account
for cluster extension, while the solid lines assume that clusters
are point sources.
The Planck--like survey is based on the Planck 2.1~mm channel
performance {\em only}: $Y=10^{-4}$~arcmin$^2$/beam and an angular
resolution of $\fwhm=5$~arcmins (this is most surely an optimistic
estimate as the Planck SZ maps will be weighted by channels with
differing resolution and sensitivity, and foreground removal will
further degrade the sensitivity).  The interferometer--like curve
assumes $\fwhm\sim 2$~arcmin and a sensitivity of 
$Y=5\times 10^{-5}$~arcmin$^2$.
}
\end{figure}

     The figure is illustrative, especially because
we have no idea of the evolution of the $L_x-T$ relation
to such redshifts.  It is nevertheless instructive to
note the different behavior of the X--ray and SZ curves, and in
particular the almost flat asymptote of the 
SZ curves; in fact, they demonstrate a slow {\em decrease} 
towards high redshift (this is independent, of course, of the
$L_x-T$ relation).  In other words, SZ selection
corresponds essentially to a mass selection.  The conclusion
depends in detail on the constancy of $\fgas$ with redshift, 
but the potential consequences are interesting:
for example, this implies that there is no $V_{\rm max}$ for
a SZ detected cluster.  Even more significant is that
SZ selection therefore finds the {\em same kinds of objects}
(mass--wise) over a large range of redshifts.  For evolutionary
studies of clusters, this is an important 
advantage over other selection techniques that
force one to compare ever more massive objects 
detected at high redshift to less massive 
local samples.

\section{Aspects of Surveying}

     The discussion so far as assumed that clusters are 
unresolved -- that the integrated SZ flux of Eq. (1)
is directly measured.  This situation applies in practice to the
Planck survey, with angular resolution of at best 5~armins.
Ground--based instruments, on the other hand, will
operate at higher resolution and clusters will
appear extended, complicating the selection criteria:
some clusters may be {\em resolved out}, relative to 
a hypothetical point--source approximation.  Detection
depends now not only on $Y$, but also on the detailed SZ
cluster profile; and photometry becomes an issue,
as with any extended source.  

\begin{figure}
\plotone{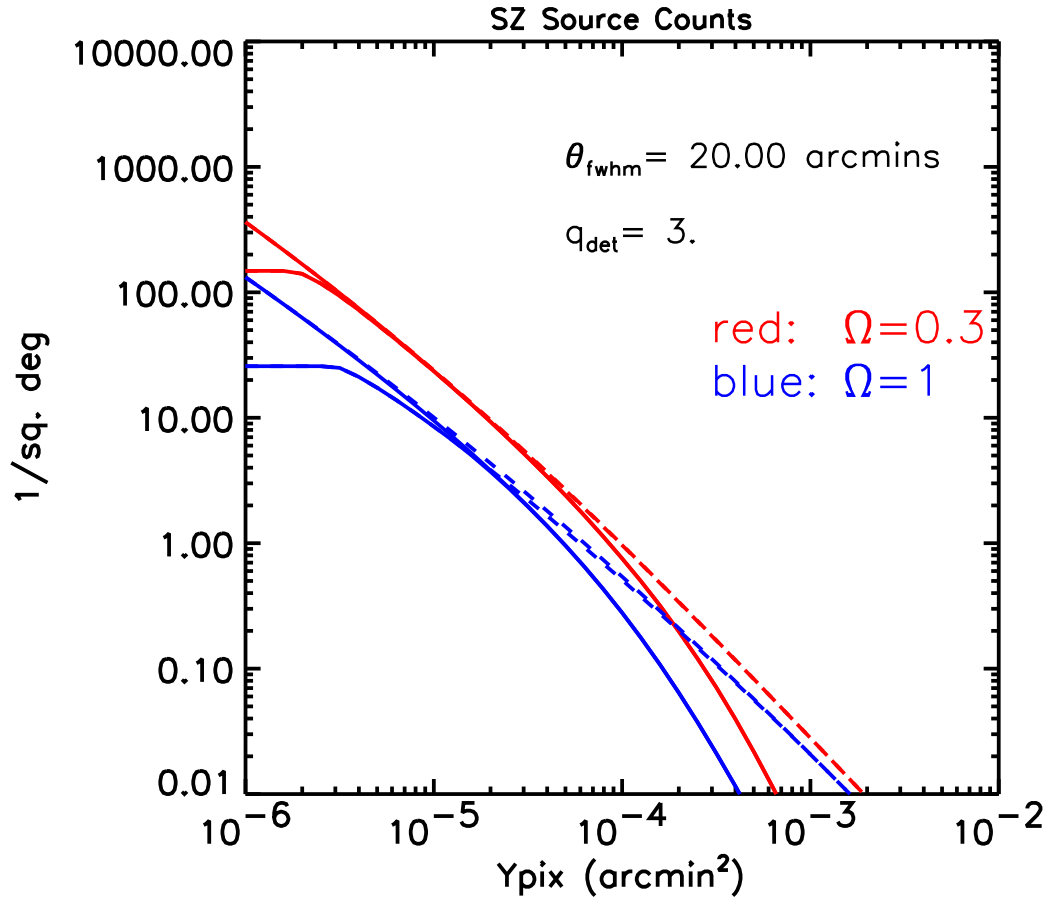}
\caption{Integrated SZ counts as a function of {\em pixel noise}
expressed in terms of $Y$ (arcmin$^2$).  Two cosmological
models are used to illustrate the effects of source extension.
The upper (red) curves are for an open model and the lower
(blue) curves are for a critical model.  The solid curves
take into account finite source extension with a simple analytical
model (Bartlett 2000), while the dashed lines
give the results of the point--source approximation.  In all
cases, two different low--mass cutoffs --
$10^{13}$ \& $10^{14}$ solar masses -- are applied in order to 
highlight the contribution of low--mass objects.  
The effect of the latter cutoff is clearly seen as the point where
the counts flatten off at low $Y$.}
\end{figure}

     Many ground--based instruments will operate, at
least at first, at a single frequency.  In this case,
primary CMB fluctuations (in the background) 
become an important source of noise, and one must
extract extended sources over a fluctuating background.
Another potentially important effect is confusion
by point sources, either randomly distributed on
the sky, or preferentially in cluster cores.  Without
spectral information, their removal requires higher resolution
imaging; for this reason, some ground--based interferometer
projects plan to control this confusion with additional
telescopes placed on long baselines.

     There are certainly other issues, for example other foregrounds,
that we could list as problematic.  However, I would instead 
like to address a specific question that often arises
when preparing a survey: In a given amount of observing time,
should one go (integrate) `deep', or instead `wide', covering
more sky to less depth?  Actually, the
real approach is to try to do several surveys, i.e., in
some sense to do both.  All the same, it is useful
to try to answer the question in terms of maximizing 
the number of detected objects.  This depends on the
slope of the source counts {\em as a function of 
survey sensititivity}\footnote{For resolved sources,
these counts are not necessarily the same as the counts
as a function of integrated cluster signal $Y$.}.

\subsection{Deep or Wide?} 

     We calculate the counts using Eq. (2).
For unresolved sources, the detection mass is only
a function of sensitivity, $\Ynoise$, and redshift -- 
$\Mdet(\Ynoise,z)$.  Additional factors are important
in the case of resolved source detection, such as angular
resolution ($\Ompix$, the solid angle of a `pixel') and
source profile: $\Mdet(\Ynoise,z,\Ompix,profile)$.
A simple analytic model of extended source 
detection in a map may be employed to examine
cluster selection (Bartlett 2000).  It supposes
that one can identify and then integrate over all
the pixels covering a cluster (out to the virial radius)
to access the total cluster signal $Y$ for detection;
it is therefore the best one can hope to do.

     Using this approach, the effects of resolution
on $\Mdet$ are presented in Figure 1, where 
the dashed lines (for the SZ curves) 
give $\Mdet(z)$ once cluster extension
has been taken into account.  Clearly,
clusters at low/intermediate redshifts are {\em resolved
out} relative to the point--source calculation, because
they are covered by several noisy pixels; this is most
evident for the higher resolution interferometer--like curve.
The difference for a low resolution survey, 
such as the Planck survey, is relatively minor, as can been
seen.

     This point is important when it comes to predicting
the number of objects to be expected.  Clearly 
the predicted number will be lower when correctly 
accounting for resolution effects.  Even more relevant
to the question at hand (deep or wide), is the fact
that, as seen in Figure 2 (and in Figure 4 below), 
the actual counts as a function
of sensitivity are steeper than the point--source
calculation would suggest.  In fact, at the bright
end they are steeper than $\Ynoise^{-2}$, the critical
slope for answering our question.  Counts steeper than $-2$ 
imply that more objects will be found by integrating deep
on a small patch of sky (as long as $\Ynoise\sim t^{-1/2}$,
which usually applies).  At the faint end, the counts
flatten out and rejoin the point--source calculation,
which is always flatter than $\Ynoise^{-2}$.
Two important conclusions follow:
\begin{enumerate}
\item the point--source
calculation gives the wrong answer in suggesting that
one should go wide; 
\item the correct answer is found
by accounting for resolution effects and indicates
that, for each experimental set--up, there is an
optimal sensitivity corresponding to the point where
the counts reach a slope of $-2$.
\end{enumerate}

\section{Expectations}

     A number of dedicated instruments optimized for SZ surveying
have recently been proposed.  At the time of writing, many of
these projects are funded and under construction, including:
\begin{itemize}
\item Interferometer arrays
\begin{itemize}
\item Arcminute MicroKelvin Imager (AMI)\\
({\tt http://www.mrao.cam.ac.uk/telescopes/ami/index.html})
\item Array for Microwave Background Anisotropy (AMiBA)\\
({\tt http://www.asiaa.sinica.edu.tw/amiba/})
\item SZ Effect Imaging Array\\ 
({\tt http://www.astro.uiuc.edu/~jmohr/SZE/index.html})
\end{itemize}
\item Bolometer cameras
\begin{itemize}
\item ACBAR\\
({\tt http://cfpa.berkeley.edu/~swlh/research/acbar.html})
\item BOLOCAM\\
({\tt http://www.astro.caltech.edu/~lgg/bolocam/bolocam.html})
\end{itemize}
\item Planck satellite\\
({\tt http://astro.estec.esa.nl/Planck/})
\end{itemize}

Several authors have recently made predictions for these various 
instruments (Aghanim et al. 1997; Holder et al. 2000; Lo et al. 2000;
Benson et al. 2001; Kay et al. 2001; 
Kneissl et al. 2001).  In Figures 3 \& 4 I show some expectations 
for the Planck Survey and for the AMI interferometer, both for 
two extreme cosmologies: a critical model and a purely open
model with $\OmM=0.3$. A flat low--density model with non--zero
cosmological constant will fall between these two cases, although
somewhat closer to the purely open model.  

      In the following,
it should be kept in mind that the predicted total number of
detected clusters is subject to large modeling uncertainties
and can vary by as much as a factor of $\sim 5$.  
A particularly important source of uncertainty comes
from the uncertainty in the amplitude of the 
density perturbations $\sigma_8$.  To be conservative, 
the predictions are given for `low $\sigma_8$': 
$(\sigma_8=0.8,\OmM=0.3)$ and $(\sigma_8=0.5,\OmM=1)$
(e.g., Eke et al. 1998).
Use of higher values quoted in the literature (e.g., Blanchard 
et al. 2000; Pierpaoli, Scott, \& White 2001) would
substantially increase the predicted total number of clusters.
Another source of uncertainty lies in the potential evolution
of the cluster gas mass fraction $\fgas$.  The lack of
evidence for strong evolution of $\fgas$ with redshift or
for an important dependence on cluster mass may simply
reflect a lack of sufficient data.  In this light, it
is important to note that the SZ signal for unresolved
clusters (e.g., Planck detections) depends {\em only} on
the global quantity $\fgas$, and not on the details
of the gas distribution; X--ray observations, as a contrast,
are heavily influenced by the physics of the core, which is
very poorly understood and difficult to model.

\begin{figure}
\plotone{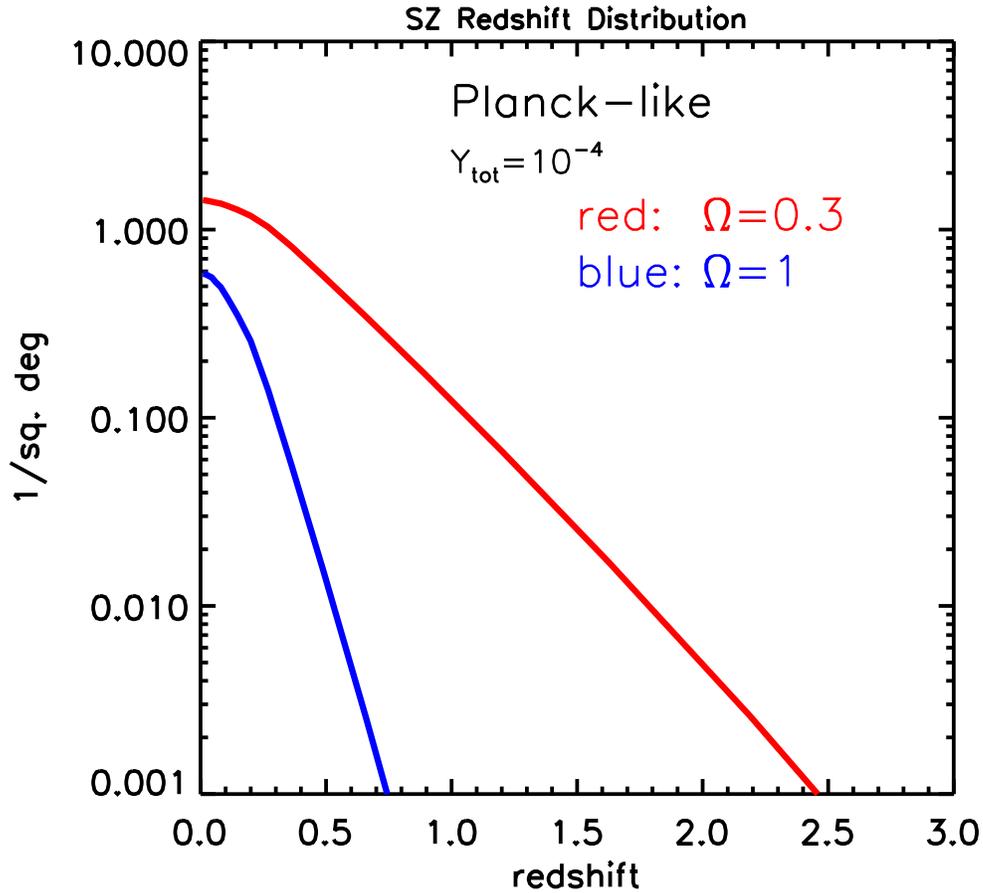}
\caption{Planck--like cumulative redshift distributions
for two extreme cosmologies.  The $y$--axis gives the counts 
at $Y>10^{-4}$~arcmin$^2$ for objects at redshifts $>z$, plotted 
as a function of $z$ ($x$--axis).  The upper (red) curve corresponds 
to the open model, while the lower (blue) curve is for the critical 
model.  In addition to the various modeling uncertainties mentioned
in the text, the predicted number of clusters
depends on the final sensitivity of the survey, which for the
fixed observing time of the mission will depend on various 
factors, including the weighting given to each of the bands,
the effects of foreground removal, etc...  This is an area
of current study.}
\end{figure}

\begin{figure}
\plotone{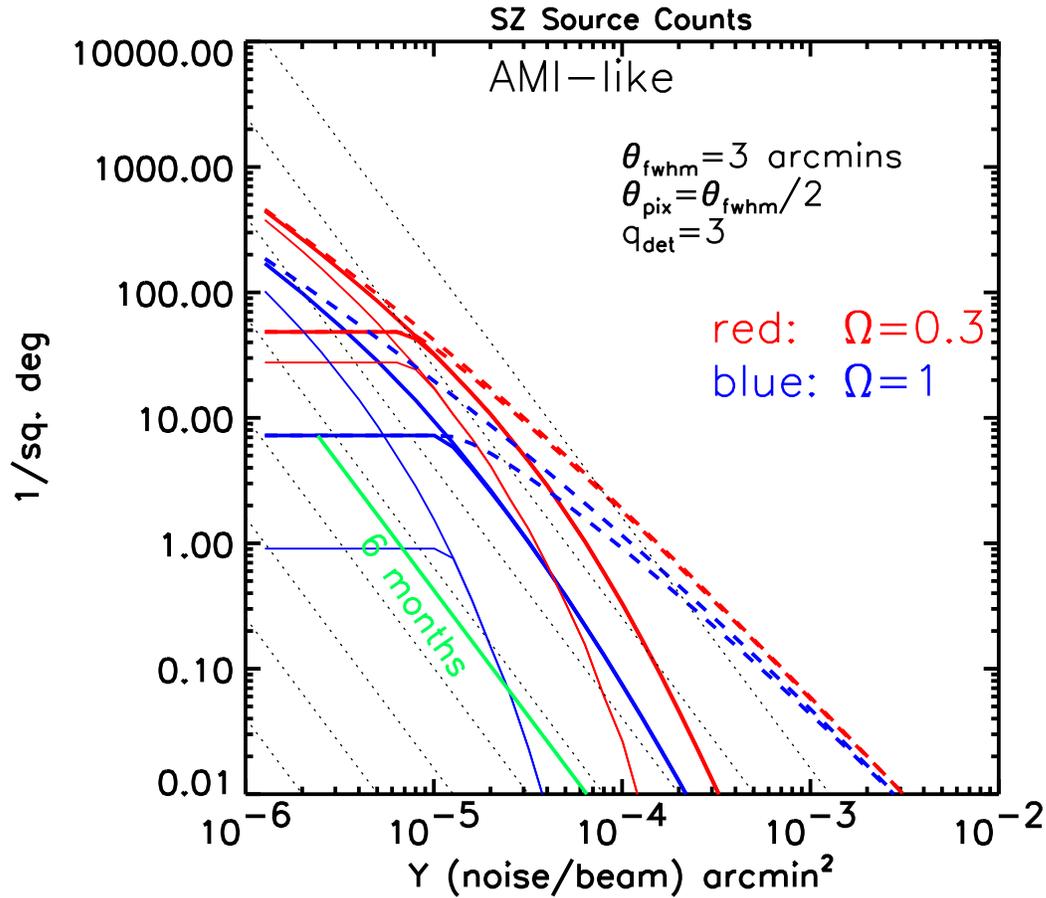}
\caption{Integrated counts for an AMI--like instrument
as a function of survey sensitivity in terms of {\em noise rms/beam}.
The color code is the same as above.  The thick solid lines
show the total counts, while the thin solid lines show only 
unresolved sources.  Each model is plotted for two different
low--mass cutoffs ($10^{13}$ and $10^{14}$ solar masses); the 
higher mass cutoff in each case can be seen to flatten off 
at the low--noise end.  The dashed
lines show the results of the (invalid) point--source 
approximation, for both mass cutoffs and for the open
model (above) and the critical model (below).  Finally,
the dotted lines in the background indicate the critical
slope of $-2$, steeper than which deep surveys yield more
objects than wide, shallow ones.}
\end{figure}

     The Planck predictions are given in terms of redshift distributions:
shown as a function of redshift $z$ are the total number of clusters 
with $Y>10^{-4}$~armin$^2$ at redshifts {\em larger} than 
$z$\footnote{i.e., the integrated counts from $z=\infty$
down to $z$, as a function of $z$ and at the fixed sensitivity
of $10^{-4}$~arcmin$^2$.}.  The numbers at $z=0$ therefore give
the total number of detectable clusters, and we see that
on the order of 1 cluster/sq.deg. is expected for the survey,
depending on the cosmology.  At $z\sim 1$ the two models differ
wildly.  There are essentially no clusters expected in the critical
model.  On the other hand, about 10\% of clusters are expected to
lie at $z>1$ in the open scenario.  
These conclusions assume that $\fgas$ remains relatively
constant with redshift and mass.  The presence of an important
population of clusters at high redshift would be a strong indication
of a low--density universe; but the lack of such a population could
be accommodated in the open model by decreasing $\fgas$ with $z$.
The result in any case would be an important piece of cosmological
information.

     The predictions for an AMI--like instrument (resolution of a few
arcminutes or better) are given in terms of integrated source counts
as a function {\em beam sensitivity}.  Both cosmological models
are once again shown, each one for two different low--mass 
cutoffs ($10^{13}$ and $10^{14}$ solar masses) in order to
separate the contribution of low and high mass objects.  
The dashed curves show the equivalent counts in a point--source 
approximation, and we observe that the latter  
fails rather badly because most clusters are resolved at 
this resolution.  The dotted lines in the background follow the 
critical slope of $Y^{-2}$.  This kind of instrument reaches
its optimal sensitivity at $\sim few\times 10^{-5}$~arcmin$^2$/beam,
where the predicted counts attain a slope of $-2$.  At this sensitivity,
one may expect, as a conservative round number and within
the modeling uncertainties mentioned,  on the order of 
several clusters/sq.deg.  

\section{Conclusion}

     It will shortly be possible to perform purely SZ--based
surveys, opening up what may be refered to as the
`SZ band' for cluster studies. A number of ground--based 
instruments will realize such surveys within a few years
over patches of sky covering several square degrees; and
by the end of the decade, the Planck mission will produce 
a catalog of several $10^4$ SZ clusters over the majority 
of the sky.  Although predictions for the number of clusters
expected for these surveys suffer from modeling 
uncertainties\footnote{Which is not surprising: if there were
no uncertainties the surveys would be of little value!},
various studies indicate roughly on the order of 
$\sim 1$~cluster/sq.deg. for Planck and $\sim few$~clusters/sq.deg. 
for the deeper ground--based efforts under way.

     These surveys represent a new method for studying clusters
that will find its place alongside X--ray and optical methods.  
As a population, clusters provide a unique statistical
sample for probing cosmology and the large--scale structure
formation process.  Each observational approach to surveying
for clusters has its proper advantages and disadvantages (collectively,
its biases).  In light of cosmological
studies, it is most instructive to quantify observational
biases in terms of cluster mass and redshift, the two primary
theoretical cluster descriptors.  
The specific advantages inherent to SZ surveying are
listed above under Section 2 and include the fact that
the integrated SZ signal is proportional to a physical quantity --
the ICM thermal energy.

     For a more complete understanding of the cluster population
in view of its application as a cosmological probe it is essential
to combine a variety of surveying (catalog selection) methods
with different biases.  SZ surveying will soon add an important
new, independent and complementary approach in this endeavor.\\

     I would like to thank the organizers for their invitation to
a most interesting and enjoyable meeting.

\end{document}